\documentclass[12pt]{article}

\usepackage{graphicx}
\usepackage{amssymb}
\usepackage{amsmath}
\usepackage{srcltx}
\usepackage{color}
\usepackage{cite}
\usepackage{url}
\usepackage{ntheorem}
\usepackage{psfrag}
\usepackage[bookmarks,colorlinks=false]{hyperref}

\theorembodyfont{\upshape}

\theoremsymbol{\ensuremath{\diamondsuit}}
\newcommand{\fcoco}{\small}
\theorembodyfont{\fcoco}\theoremseparator{}\theoremindent0.5cm

\theoremstyle{nonumberplain}\theorembodyfont{\fcoco}
\theoremseparator{}\theoremindent0.5cm

\DeclareFontFamily{OT1}{rsfs}{}
\DeclareFontShape{OT1}{rsfs}{m}{n}{ <-7> rsfs5 <7-10> rsfs7 <10-> rsfs10}{}
\DeclareMathAlphabet{\mycal}{OT1}{rsfs}{m}{n}

{\catcode `\@=11 \global\let\AddToReset=\@addtoreset}
\AddToReset{equation}{section}

\newcounter{mnotecount}[section]

\AddToReset{figure}{section}
\renewcommand{\themnotecount}{\thesection.\arabic{mnotecount}}

\newcommand{\mnotex}[1]
{\protect{\stepcounter{mnotecount}}$^{\mbox{\footnotesize
$
\bullet$\themnotecount}}$ \marginpar{
\raggedright\tiny\em
$\!\!\!\!\!\!\,\bullet$\themnotecount: #1} }

\newcommand{\ptccheck}[1]{\mnote{ptcchecked:#1}}
\renewcommand{\ptccheck}[1]{\mnote{\checkmark (ptc #1)}}

\newcommand{\jlcax}[1]{}
\newcommand{\eean}{\nonumber\end{eqnarray}}

\newcommand{\kk}[1]{}

\newcommand{\beq}{\begin{equation}}

\newcommand{\FS}       
                  {F}

\newcommand{\HS} 
       {H_{\mbox{\scriptsize volume}}}

{\ptc{this should be removed in the oberwolfach version}}%

\newcommand{\eeal}[1]{\label{#1}\end{eqnarray}}
\newcommand{\bed}{\begin{deqarr}}
\newcommand{\eed}{\end{deqarr}}
\newcommand{\bedl}[1]{\begin{deqarr}\label{#1}}
\newcommand{\eedl}[2]{\arrlabel{#1}\label{#2}\end{deqarr}}

\newcommand{\bel}[1]{\begin{equation}\label{#1}}
\newcommand{\bea}{\begin{eqnarray}}
\newcommand{\bean}{\begin{eqnarray}\nonumber}
\newcommand{\beal}[1]{\begin{eqnarray}\label{#1}}
\newcommand{\eea}{\end{eqnarray}}

\newcommand{\be}{\begin{equation}}
\newcommand{\eeq}{\end{equation}}
\newcommand{\ee}{\end{equation}}
\newcommand{\beqa}{\begin{eqnarray}}
\newcommand{\eeqa}{\end{eqnarray}}
\newcommand{\beqan}{\begin{eqnarray*}}
\newcommand{\eeqan}{\end{eqnarray*}}
\newcommand{\ba}{\begin{array}}
\newcommand{\ea}{\end{array}}

\newcommand{\mnote}[1]
{\protect{\stepcounter{mnotecount}}$^{\mbox{\footnotesize
$
\bullet$\themnotecount}}$ \marginpar{
\raggedright\tiny\em
$\!\!\!\!\!\!\,\bullet$\themnotecount: #1} }

\newcommand{\warn}[1]
{\protect{\stepcounter{mnotecount}}$^{\mbox{\footnotesize
$
\bullet$\themnotecount}}$ \marginpar{
\raggedright\tiny\em
$\!\!\!\!\!\!\,\bullet$\themnotecount: {\bf Warning:} #1} }

\newcommand{\Z}{\mathbb Z}

\newcommand{\ptc}[1]{\mnote{{\bf ptc:}#1}}

\newcommand{\beqar}{\begin{deqarr}}
\newcommand{\eeqar}{\end{deqarr}}

\newcommand{\beaa}{\begin{eqnarray*}}
\newcommand{\eeaa}{\end{eqnarray*}}

\renewcommand{\ptccheck}[1]{}
\begin{document}
\title{Stable causality of the Pomeransky-Senkov black holes}

\author{Piotr T.~Chru\'sciel\thanks{PTC was supported in part  by the EC project
KRAGEOMP-MTKD-CT-2006-042360, and by the Polish Ministry of
Science and Higher Education grant Nr N N201 372736.}
\\
Faculty of Physics, University of Vienna
\\
\\
Sebastian J. Szybka\thanks{SSz was supported in part by the
Polish Ministry of Science and Higher Education grant Nr N N202
079235,
and by the Foundation for Polish Science.} \\
Obserwatorium Astronomiczne, Uniwersytet Jagiello\'nski,
Krak\'ow}
\date{October 1, 2010}

\maketitle{}
\begin{abstract}
We show stable causality of the Pomeransky-Senkov black rings.
\end{abstract}

\section{Introduction}
\label{s30IX0.2}

Five-dimensional black rings, and their generalisations, have
attracted a lot of attention in recent literature (see, e.g.,
\cite{EmparanReallLR}). In~\cite{CCGP} it has been shown that
the Pomeransky-Senkov~\cite{PS} metrics, with appropriate
values of parameters, do not contain naked singularities in
their domains of outer communications (d.o.c.). In that
reference the question of causality violations within the
d.o.c.\ has been left open, except for reporting some numerical
evidence. The object of this note is to point out that the
Pomeransky-Senkov (PS) black holes are stably causal.

Ideally one would like to show that the d.o.c.'s of the PS
metrics are globally hyperbolic, but such a result lies outside
of the scope of this work.

\section{Stable causality}

We use the conventions and notations of~\cite{CCGP}, except
that we write $G(x,\lambda,\nu)$ for $G(x)$ from~\cite{CCGP},
etc.
In that reference it has been shown that
\bean
 g(\nabla t, \nabla t) = g^{tt} &=& \frac{g_{xx}
 g_{yy}}{\det g_{\mu \nu}}\det
 \left( \begin{array}{cc}
          g_{\psi \psi} & g_{\psi \varphi} \\
          g_{\psi \varphi} & g_{\varphi \varphi}
        \end{array}
        \right)
\\
 & = &\frac{   (\nu -1)^2
(x-y)^4}{ 4  k^4 G(x,\lambda,\nu) G(y,\lambda,\nu)   }\det
 \left( \begin{array}{cc}
          g_{\psi \psi} & g_{\psi \varphi} \\
          g_{\psi \varphi} & g_{\varphi \varphi}
        \end{array}
        \right)\nonumber\\
&=:&\frac{(1+y)(1-x^2)\Theta(x,y,\lambda,\nu)}{(1-\lambda+\nu)H(x,y,\lambda,\nu)G(x,\lambda,\nu)G(y,\lambda,\nu)}
 \;,
\eeal{7IX.2}
where $\Theta$ is a \emph{polynomial} in the coordinates $x$,
$y$, and in the parameters $\lambda$ and $\nu$, whose exact
form it too complicated to be usefully displayed
here. On the
d.o.c.\ of the PS metrics we have
$$
 x\in [-1,1]\;, \quad
  y \in (y_h, -1]\;,
   \quad
   \nu\in (0,1)\;,\
    2\sqrt{\nu}\le \lambda< 1 +\nu
    \;,
$$
where
$$
y_h:= -\frac{\lambda-\sqrt{\lambda^2-4\nu}}{2\nu} >
-\frac{\lambda+\sqrt{\lambda^2-4\nu}}{2\nu} =:y_c\;.
$$
Stable causality of the d.o.c. will follow if one can prove that
$g^{tt}$ is strictly negative there.
Away from the boundaries $y=-1$ and
$x=\pm 1$, this is equivalent to strict negativity of $\Theta$.
This remains true on those boundaries because
$$
 G(y,\lambda,\nu)=\left(1-y^2\right) \left(\nu y^2+\lambda  y+1\right)
 \;.
$$
This shows that the multiplicative factor $(1+y)$ in the
numerator of $g^{tt}$ is cancelled by the first order zero of
$G(y,\lambda,\nu)$, so $\nabla t$ will again be timelike at $y=-1$ if
$\Theta$ is strictly negative there. An identical argument
applies to $x=\pm 1$.

The following change of variables
can be used to show that $\Theta$ has a sign: let $a\in
[0,\infty)$ and $d\in (0,\infty)$, the redefinition
$$
 x = -1 + \frac 2 {1+a}\;,
  \quad
  \nu = \frac 1{(1 + d)^2}\;,
$$
leads to the right ranges of $x$ and $\nu$, except for $x=-1$
which will be considered later. Setting
$$
  \lambda  = 2 \frac{2 d^2 + 2 (2 + c) d+ (2 + c)^2}{(2 + c) (1 + d) (2 +
c + 2 d )}\;,
$$
where $c\in (0,\infty)$ covers the range of allowed
$\lambda$'s, except for the borderline case $\lambda =2 \sqrt
\nu$ (to which we will return shortly); to check this it is
useful to note that
$$
 \partial _ c \lambda = -
 \frac{8d^2 (2+c+d)}{(2+c)^2(1+d)(2+c+2d)^2} < 0
 \;.
$$
Finally, the formula
$$
 y =  -1 - \frac{d (4 + c+2 d )}{(1 + b) (2 + c)}
$$
leads to $y$ in the   range  $[y_c,-1)$ if $b\in [0,\infty)$,
which is more than needed for $(y_h,-1)$, but note that $y=-1$
is missing.

Inserting the above into $\Theta$ one obtains a rational
function with denominator
$$
 (1+a)^4 (1+b)^5 (2 + c) ^9  (1+d)^{12}(2+c+2d)^5
 \;.
$$
and with numerator which is a polynomial, say $P$, in
$(a,b,c,d)$. A {\sc Mathematica} calculation shows that all
coefficients are \emph{negative integers} in
$$[-61382522306560 ,-1] \cap \Z
 \;.
$$
One also finds
$$
 P <  -2048 c^{4} d^{23}
 \;,
$$
which proves \emph{strict} negativity of $\Theta$ away from the
boundaries $x=-1$ and $y=-1$, for non-extreme configurations
$\lambda > 2\sqrt \nu$.

Consider now the case $y=-1$.  We proceed as before, except
that we first set $y=-1$ in $\Theta$, and then replace
$(x,\lambda,\nu)$  by $(a,c,d)$. The end result is a rational
function with denominator
$$
 (1+a)^4 (2 + c) ^5 (1+d)^{12}  (2+c+2d)^5
 \;,
$$
with a numerator, say $R$, a polynomial with strictly
negative coefficients belonging to
$$
 [ - 19763036160,-2] \cap \Z
 \;,
$$
satisfying
$$
 R < -2 c^{10} d^{12}
 \;,
$$
hence strictly negative.

The case $x=-1$ is analysed in a similar way.

When $\lambda=2\sqrt \nu$ strict negativity of $\Theta$ is
established by using instead
$$
 y =  -1 - \frac{d  }{(1 + b)  }
$$
in the arguments above.

\bigskip

\noindent{\sc Acknowledgements} We are grateful to Alfonso
Garcia-Parrado for making his {\sc
Mathematica-xAct}~\cite{xAct} files available to us.

\bibliographystyle{amsplain}
\bibliography{../references/hip_bib,%
../references/reffile,%
../references/newbiblio,%
../references/newbiblio2,%
../references/bibl,%
../references/howard,%
../references/bartnik,%
../references/myGR,%
../references/newbib,%
../references/Energy,%
../references/netbiblio}
\end{document}